\documentclass[11pt,preprint]{aastex}
\usepackage{graphicx}
\usepackage{epsfig}
\usepackage{subfigure}

\let\oldfootsep=\footnotesep
\newcommand\ltsima{$\; \buildrel <\over\sim \;$}
\newcommand\simlt{\lower.5ex\hbox{\ltsima}}
\newcommand\gtsima{$\; \buildrel >\over\sim \;$}
\newcommand\simgt{\lower.5ex\hbox{\gtsima}}
\newcommand\etal{et~al.}

\newcommand\msun {M_\odot}

\newcommand\mearth {M_\oplus}

\newcommand\Amax{A_{\rm max}}

\newcommand\pac{Paczy{\'n}ski }
\newcommand\ie{{\it i.e. }}

\newcommand\rep {\tilde{r}_e}
\newcommand{\mathbold}[1]{\mbox{\boldmath $\bf#1$}}

\newcommand\be{\begin{equation}}
\newcommand\ee{\end{equation}}

\def\s#1{\sigma_{#1}}
 
\shorttitle{}
\shortauthors{Bennett, Anderson \& Gaudi}

\begin{document}


\title{Characterization of 
       Gravitational Microlensing Planetary Host Stars}


\author{David~P.~Bennett\altaffilmark{1},
               Jay Anderson\altaffilmark{2}, and
               B.~Scott Gaudi\altaffilmark{3}
       } 
\altaffiltext{1}{Department of Physics,
    University of Notre Dame, IN 46556, USA\\
    Email: {\tt bennett@nd.edu}}

\altaffiltext{2}{Astronomy Department,
    Rice University, Houston, TX, USA\\
    Email: {\tt jay@eeyore.rice.edu} }

\altaffiltext{3}{Department of Astronomy, 
                        The Ohio State University, 
                        140 W. 18th Ave., 
                         Columbus, OH 43210\\
    Email: {\tt gaudi@astronomy.ohio-state.edu} }


\clearpage

\begin{abstract}
The gravitational microlensing light curves that reveal the presence
of extrasolar planets generally yield the planet-star mass ratio and
separation in units of the Einstein ring radius. The microlensing
method does not require the detection of light from the planetary 
host star. This allows the detection of
planets orbiting very faint stars, but it also makes it difficult to 
convert the planet-star mass ratio to a value
for the planet mass. We show that in many cases, the lens
stars are readily detectable with high resolution space-based 
follow-up observations in a single passband.
When the lens star is detected, the
lens-source relative proper motion can also be measured, and this
allows the masses of the planet and its host star to be determined and
the star-planet separation can be converted to physical units.
Observations in multiple passbands provide redundant information,
which can be used to confirm this interpretation.
For the recently detected super-Earth
planet, OGLE-2005-BLG-169Lb, we show that the lens star
will definitely be detectable with observations by the
Hubble Space Telescope (HST) unless it is a stellar remnant.
Finally, we show that most planets detected by a space-based
microlensing survey are likely to orbit host stars that will be
detected and characterized by the same survey.
\end{abstract}




\section{Introduction}
One of the primary strengths of the gravitational microlensing extrasolar
planet detection method is its sensitivity to low-mass
planets at separations of a few AU. This has recently been demonstrated 
with the discovery of two of the lowest mass extrasolar planets known:
OGLE-2005-BLG-390Lb, a $5.5{+5.5\atop -2.7} \mearth$ planet orbiting an
M-dwarf in the Galactic bulge \citep{ogle390}, and OGLE-2005-BLG-169Lb, 
a $13 {+6\atop -8} \mearth$ which probably orbits a K or early M star in the inner 
Galactic disk \citep{ogle169}. Both of these planets have a separation
from their host star of about $\sim 3\,$AU, and this places them well
outside the current sensitivity range of the radial velocity method, which
is only sensitive to such planets with a separation of $\simlt 0.7\,$AU
\citep{rivera_gj876d,lovis_3nep}.

The microlensing method is
most sensitive to planets at a separation of 1-5$\,$AU, and this separation
region is particularly interesting from a theoretical point of view because
it contains the so-called ``snow-line" which is an important feature for
the core accretion theory of planet formation. This ``snow-line" is
the region of the proto-planetary disk where it is cold enough for
water-ice to condense, and the core accretion theory predicts that
this is where the most massive planets will form 
\citep{ida_lin,laughlin,kennedy-searth}. According to this
theory, giant planets form just outside the ``snow line" where they
can accrete $\sim 10\mearth$ of rock and ice to form a core
that grows into a gas giant like Jupiter or Saturn via the run-away 
accretion of Hydrogen and Helium onto this core.
However, this theory also predicts that the Hydrogen and Helium
gas can easily be removed from the proto-planetary disk during the
millions of years that it takes to build the rock-ice core of a gas-giant.
Thus, if the core accretion theory is correct, rock-ice planets of 
$\sim 10\mearth$ that failed to grow into gas giants
should be quite common. Recent core-accretion theory
calculations \citep{laughlin,ida_lin,boss-giant} predict that it is especially
difficult to form gas-giant planets around low-mass stars. 

As of November, 2006, there have been four extrasolar planets discovered
by the microlensing method: two planets more massive than Jupiter
\citep{bond-moa53,ogle71}  in addition to the two $\sim 10\mearth$
super-Earth 
planets. But, the microlensing planet detection efficiency is more than
an order of magnitude larger for Jupiter-mass planets than for 
$10\mearth$ planets, so the microlensing results indicate that 
planets of $\sim 10\mearth$ are significantly more common than
Jupiters \citep{ogle390} and that 16-69\% (90\% c.l.) of
non-binary stars have a planet of $\sim 10\mearth$ \citep{ogle169}
at a separation of $\sim 1.5$-4$\,$AU.
The discovery of these two planets suggests that 
super-Earth planets of $\sim 10\mearth$
are more common than Jupiter mass planets at separations of a few
AU around the most common stars in our Galaxy. This would seem to 
confirm a key prediction of the core accretion model for planet
formation: that Jupiter mass planets are much more likely to
form in orbit around G and K stars than around M stars \citep{laughlin,ida_lin},
although super-Earth planets are also predicted in the gravitational
instability model \citep{boss-supearth}.
In fact, \citep{laughlin} have argued that the Jupiter-mass planets found
in microlensing events \citep{bond-moa53,ogle71} are more likely to orbit 
white dwarfs than M dwarfs. Clearly, the microlensing detections would
provide tighter constraints on the theories if the properties of the host
stars were known.

In this paper, we show that the planetary host stars can often be identified
in high resolution images
taken a few years after the microlensing event. In cases
where the lens star has been identified, we show in \S~\ref{sec-pm} that
a complete solution to the microlensing event can generally be found. 
The complete solution gives the lens star and
planet masses and the separation in physical units instead of just the
star-planet mass ratio and separation in Einstein ring radius units. 
This has recently been demonstrated with the first planet discovered
by microlensing, OGLE-2003-BLG-235Lb/MOA-2003-BLG-53Lb
\citep{bennett-moa53}, and in \S~\ref{sec-ob169},
we consider the more favorable case of the low-mass planetary
microlensing event OGLE-2005-BLG-169, where the lens star will be 
detectable if it is a main sequence star of any mass. In \S~\ref{sec-mpf},
we show that a space-based microlensing survey \citep{gest-sim},
like the proposed Microlensing Planet Finder (MPF) mission 
\citep{mpf-spie} will detect the planetary host stars for most of the
detected planets, and so, planets detected by MPF will usually
come with a complete solution of the lens system. We conclude,
in \S~\ref{sec-conclude}, with a discussion of follow-up observations
of planetary microlensing events and argue that the host stars will
generally be detectable unless they are stellar remnants or brown
dwarfs.

\section{Lens-Source Relative Proper Motion and Complete Solution of Microlensing Events}
\label{sec-pm}

Gravitational microlensing can be used to study objects, such as brown dwarfs,
stellar remnants, or extrasolar planets that emit very little detectable radiation,
but most microlensing events provide only a single parameter, the Einstein
radius crossing time, $t_E$, that can constrain the lens system mass, distance,
and velocity. The situation is significantly improved when the lens-source
relative proper motion, $\mu_{\rm rel}$, can be determined, because 
this yields the angular Einstein radius: $\theta_E = \mu_{\rm rel} t_E$.
The angular Einstein radius is related to the lens system mass by
\begin{equation}
M_L = {c^2\over 4G} \theta_E^2 {D_S D_L\over D_S - D_L} \ ,
\label{eq-m_thetaE}
\end{equation}
where $D_L$ and $D_S$ are the lens and source distances, respectively. 
$M_L = M_\ast + M_p$ is the total lens system mass: the sum of the star
and planet masses. Since
$D_S$ is known (at least approximately), eq.~\ref{eq-m_thetaE} can be considered
to be a mass-distance relation for the lens star.

\begin{figure}
\plotone{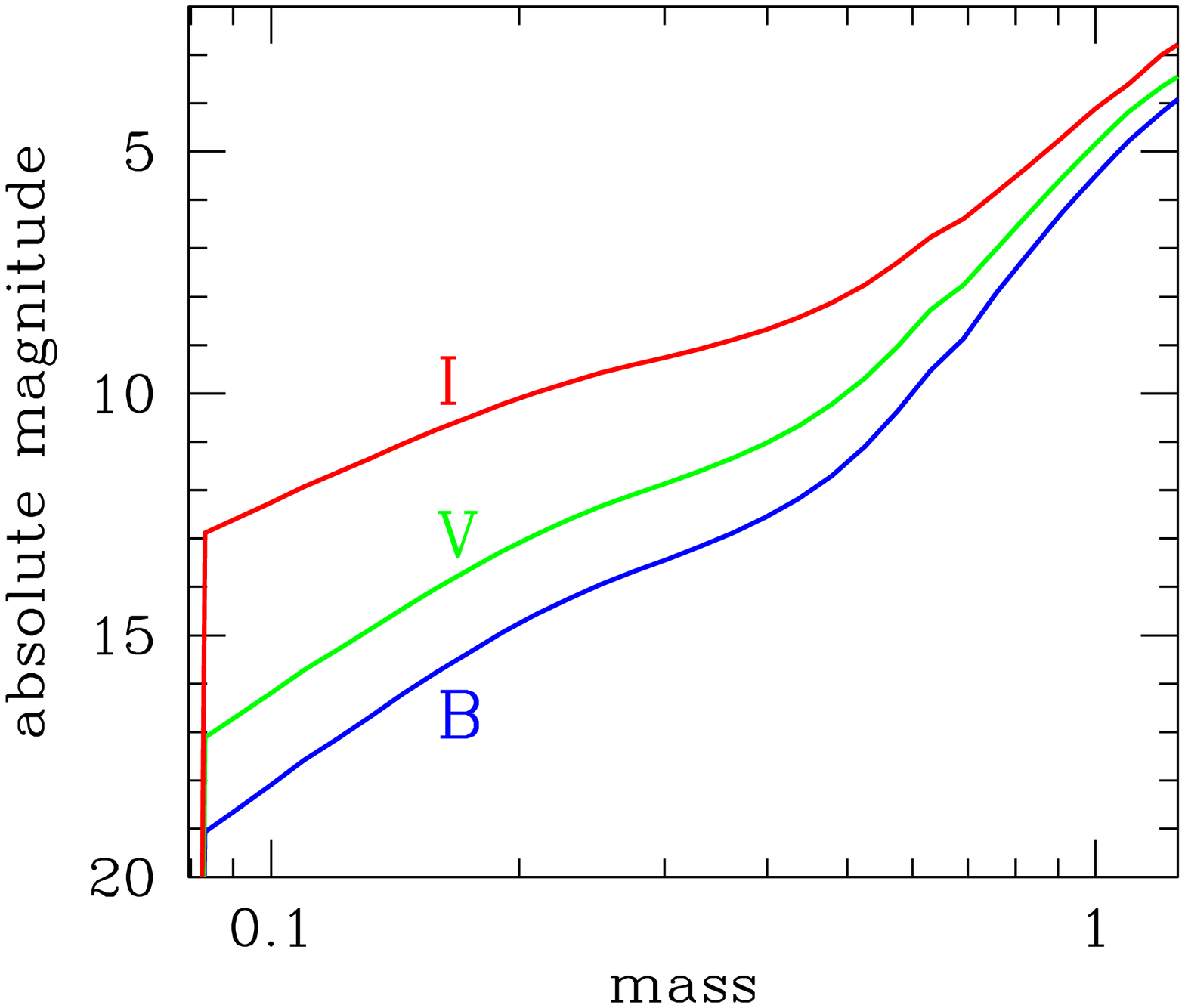}
\caption{
The adopted mass-luminosity relations in the I (F814W), V (F555W), and B (F435W)
passbands. These an empirical mass-luminosity relations
\citep{kroupa_tout} extended to higher masses and the $B$-band
using \citet{allen}, \citet{schmidt-kaler} and \citet{gray}. Approximately Solar
metallically is appropriate for the lens stars in the inner Galactic disk and the 
bulge.
\label{fig-mass_lum}}
\end{figure}

Another constraint on the lens system is needed in order to convert
this mass-distance relation to a complete solution of the lens event.
For most events, this can be accomplished by direct detection 
of the planetary host star. If the brightness of the host (lens) star is
measured, then the
mass-distance relation, eq.~\ref{eq-m_thetaE}, can be combined 
with a mass-luminosity relation, such as those shown in 
Figure~\ref{fig-mass_lum}, to yield the mass of the lens system. 
This then allows the individual star and planet masses to be determined,
since the mass ratio is known from the microlensing light curve. This
also allows the separation determined from the light curve to be
converted from units of the Einstein radius, $R_E$, to physical units.

There is still some residual uncertainty due to the uncertainty in
$D_S$, but this is generally small as can be illustrated with 
some special cases. If $D_L \ll D_S$, then eq.~\ref{eq-m_thetaE}
indicates that $M_L$ becomes independent of $D_S$. In the opposite
extreme, with the lens very close to the source ($D_L \approx D_S$),
we can treat $D_S$ and $D_S-D_L$ as the independent variables
and replace $D_L$ by $D_S$ (except in the expression $D_S-D_L$),
so in this limit $M_L \propto D_S^2/(D_S-D_L)$
The mass luminosity relations shown in Figure~\ref{fig-mass_lum}
can be approximated by a power law over a limited range of masses.
So, if we take
${\cal L} \propto M^n$ as our mass-luminosity relation, the lens
mass is simply given by $M_L \propto F_L^{1/n} D_S^{2/n}$,
where $F_L$ is the measured flux of the lens star. 
(Eq.~\ref{eq-m_thetaE} then gives $D_S-D_L$ in terms of $M_L$.)
Since the appropriate value of $n$ is almost always $n > 2$, the
fractional error in $M_L$ due to the uncertainty in $D_S$ will 
generally be smaller than the fractional uncertainty in $D_S$.

It is important to note that this complete solution of the lens system
requires only the lens brightness in a single pass-band. Color information
obtained by multi-band observations is redundant, and can be used to
constrain the possibility that the source or lens star has a bright
binary companion.

An alternative method for the complete solution of a microlensing event is to
combine a measurement of the lens-source proper motion with a measurement
of the microlensing parallax effect. This
has already been demonstrated for microlensing events due to single
stars \citep{dck-lmc5,lmc5-mass} and stellar binaries \citep{planet-er2000b5}.
The detection of the microlensing parallax effect provides
a measurement of $\rep$, the projected Einstein radius
(projected from the position of the source to that of the observer).
As originally shown by \citet{gould-par1}, the measurement of both
$\theta_E$ and $\rep$ yields the lens system mass,
\begin{equation}
M_L = {c^2\over 4G} \rep \theta_E \ ,
\label{eq-m_rep}
\end{equation}
without any dependence on $D_S$ or $D_L$. However, $\rep$
is generally measured by detecting the effect of the Earth's
orbital motion in the light curve \citep{macho-par1}, and this means that
reliable measurements of $\rep$ can generally only be made
for very long duration events or events where the lens-source
motion is unusually small 
\citep{lmc5-detect,dck-lmc5,lmc5-mass,planet-er2000b5,
mao-par,ogle-par2,moa-par,bennett-parbh,multi-par}.
For shorter events, the light curve may reveal only a single 
component of the two-dimensional ${\bf \rep}$ vector. It is then possible
to determine the lens mass from eq.~\ref{eq-m_rep} if the direction
of ${\mathbold \mu_{\rm rel}}$ is measured, since these two vectors
are parallel.

\subsection{Lens-Source Relative Proper Motion Determination Methods}

There are two methods of measuring the relative proper motion.  The
first involves a measurement of finite source effects in the microlensing
light curve combined with an estimate of the angular size of the source.
These finite source effects are commonly detectable in binary lens
events \citep{mao-pac}, when the angular position of the source crosses or comes
very close to the caustic curve of the lens
\citep{macho-lmc9,joint-98smc1,macho-binaries,ogle-bin_lens_mass}. 
Finite source effects
are particularly common in planetary microlensing events because the
planet is likely to be detected only if the source crosses or closely
approaches the caustic curve. In fact, all four planetary microlensing
events observed to date \citep{bond-moa53,ogle71,ogle390,ogle169}
have revealed finite source effects, which reveal 
the source crossing time, $t_\ast$. The angular radius
of the source star, $\theta_\ast$, can be estimated from its color and 
brightness. This yields the relative proper motion, 
$\mu_{\rm rel} = \theta_\ast/t_\ast$, and the angular Einstein radius,
$\theta_E = \theta_\ast t_E/t_\ast$, because $t_E$ can virtually
always be determined from the light curve of planetary microlensing
events.

Even if the light curve does not reveal any finite source effects, it is still
possible to measure $\mu_{\rm rel}$ directly by detecting the
lens star and measuring the lens-source separation as was done
for microlensing event MACHO-LMC-5 \citep{lmc5-detect}.
For this event, the relative proper motion was quite large,
$\mu_{\rm rel} = 21.4\,$mas/yr, because the
distance to the lens is small, $D_L = 550\,$pc. As a result, the lens
and source were easily resolved in HST images taken 6.3 years after
peak magnification.

Galactic bulge
microlensing events with planetary signals generally have relative
proper motions that are much smaller than this, typically,
$\mu_{\rm rel}  \sim 5\,$mas/yr. Thus, the lens stars would typically
not be resolved in HST images taken less than a decade after
peak magnification. But, due to the stability of the HST point-spread
function (PSF), it is possible to measure lens-source separations that
are much smaller than the width of the PSF. This is accomplished by
measuring the elongation of the combined lens-source
image due to the fact that it consists of two point sources instead
of one. 

\subsection{Image Elongation}
\label{sec-distort}

\begin{figure}
\plotone{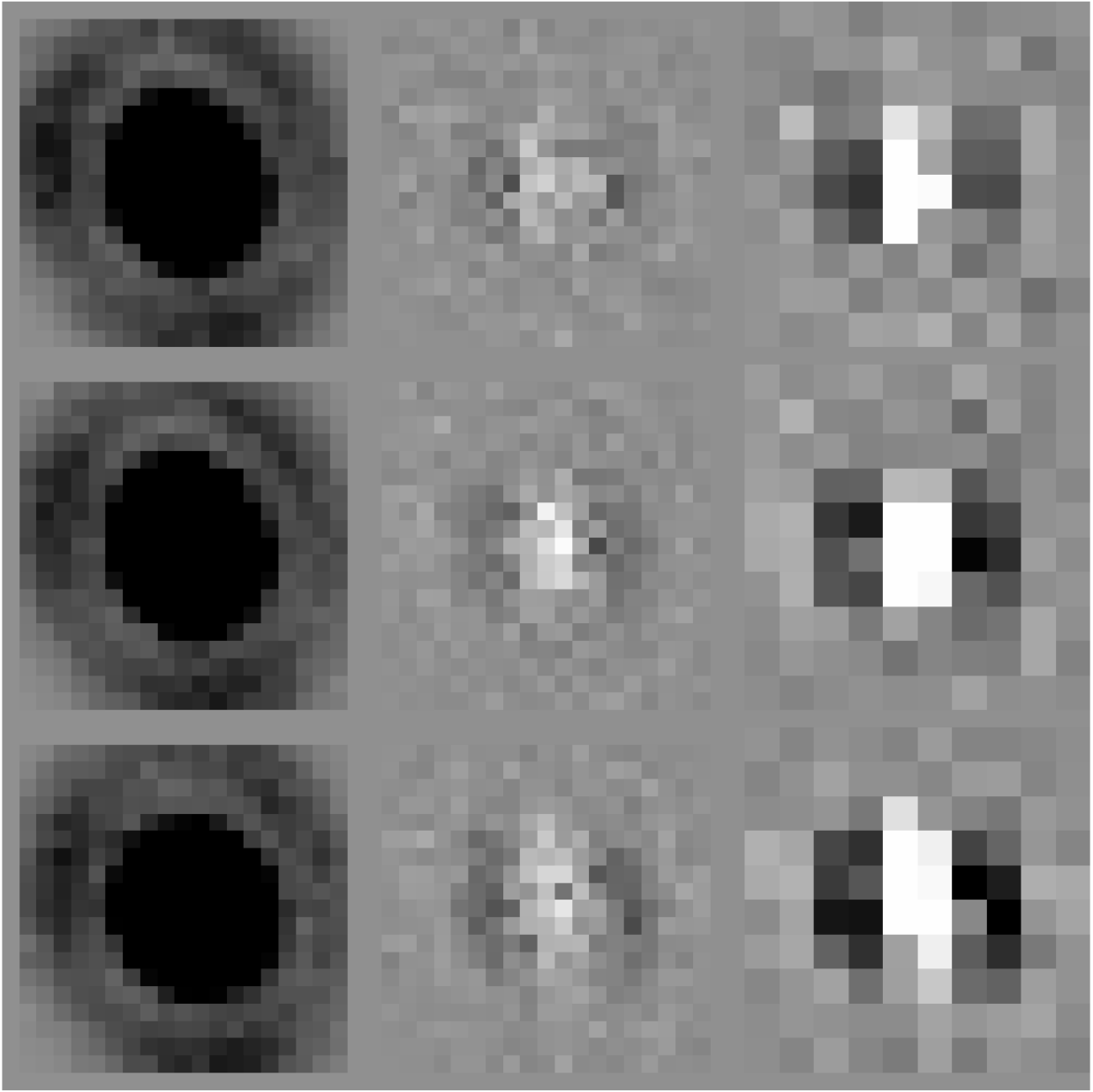}
\caption{
Simulated image stacks of multiple dithered exposures of the 
OGLE-2005-BLG-169 source and lens star 2.4 years after peak magnification
using the HST/ACS High Resolution Camera (HRC) in the F814W filter 
band. The top row of images assumed a host star mass of 
$M_\ast = 0.08\msun$, the middle row assumes $M_\ast = 0.35\msun$, and 
the bottom row assumes $M_\ast = 0.63\msun$. In each row, the image 
on the left shows the raw image stack sampled at one half the native 
HRC ($28\,$mas) pixel size. The central column shows the residuals after subtraction 
of the best fit PSF model, showing the blended image elongation along 
the $x$-axis due to the lens-source 
separation. The right hand column shows these residuals
rebinned to the $28\,$mas native pixel scale.
\label{fig-pics}}
\end{figure}

Figure~\ref{fig-pics} shows a simulated images from the
Hubble Space Telescope (HST) Advanced Camera for Surveys
(ACS) of the source and lens star for microlensing event 
OGLE-2006-BLG-169 taken 2.4 years after peak magnification
when the lens-source separation is predicted to be 
$20.2\pm 1.7\,$mas. These simulated images represent the
co-added dithered exposures from a single orbit of observations
in the $I$-band (F814W), and the three different rows of images
represent three possible masses for the planetary host star.
While the $20.2\pm 1.7\,$mas lens-source separation is
smaller than the diffraction limited $I$-band
PSF FWHM of $\sim 76\,$mas,
the elongation of the PSF of the blended source-lens star pair
is clearly visible in the residual images shown in the 
central and right hand columns. 

The high precision image elongation measurements that we
desire depend on a precise knowledge of the PSF, but
this is to be expected with space-based observations
of microlens source stars. The Galactic bulge fields where
microlensing events are discovered have large numbers
of relatively bright, but reasonably well separated 
stars in the field of the lens, and this enables the
determination of very accurate PSF models, as long
as the space telescope observations are dithered
to correct for image undersampling. Prior HST programs 
have demonstrated that HST has the requisite
image stability \citep{lauer99,andking00,andking04}.

There is a potential degeneracy encountered when 
converting the image elongation into a value for the lens-source 
displacement. The elongated image full-width half-maximum (FWHM)
in the direction of the lens-source separation is determined by 
the lens-source separation and their brightness ratio. Thus, if we
only measure the increased FWHM of the elongated image,
we won't be able to determine the brightness ratio and the 
separation. Fortunately, the microlensing 
light curves with definitive
planetary signals will generally provide the information needed to
break this degeneracy. These light curves have sufficient high precision
photometry to determine the brightness of the source star from the
shape of the microlensing light curve. Furthermore, most planetary
light curves also exhibit finite source effects, which allow an
independent determination of the lens-source relative proper motion,
$\mu_{\rm rel}$. For
these events, we can consider the image elongation to be
a measure of the the brightness ratio, which can be compared
to the source star brightness from the light curve as a test for a
bright binary companion to the source star.

The most reliable way to estimate the precision of our measurement
of $\mu_{\rm rel}$ is with Monte Carlo simulations using 
realistic PSF models, which are well understood in the case
of HST. However, it is instructive two consider a much
simpler procedure. Let $\Delta x$ be the lens-source separation
at a time, $T$, after peak magnification. The relative proper motion
is therefore $\mu_{\rm rel} = \Delta x/T$.
If we approximate the PSF with a 
Gaussian profile, $e^{-0.5x^2/s_0^2}$, then the blended image 
of the source plus lens star will be represented by the sum of 
two Gaussian profiles with centers separated by $\Delta x$.
We can then approximate this sum of Gaussians with a broader
Gaussian, $e^{-0.5x^2/s^2}$, where we require that the
broader Gaussian has the same RMS value as the original
sum of Gaussians (in the direction of the 
lens-source proper motion). This yields
\be
s^2=s_0^2 + f_L(1-f_L)(\Delta x)^2 \ ,
\label{eq-s}
\ee
where $f_L=F_L/F_{\rm tot}$ is 
the ratio of the flux of the lens to the total
flux in the PSF. Since 
the source flux $F_S = F_{\rm tot} - F_L$ is known
from the light curve, and $F_{\rm tot}$ can be measured from
the images being analyzed, we can consider
$f_L$ to be known from the light curve.  The
uncertainties in the lens-source separation, $\Delta x$,
and relative proper motion, $\mu_{\rm rel}$  can be derived,
\be\label{eq-smu_asym}
\frac{\s{\Delta x}}{\Delta x} =
\frac{\s{\mu}}{\mu_{\rm rel}} = 
\frac{1}{\sqrt{2N_{tot}}}
\left(\frac{s_0^2}{f_L(1-f_L)(\Delta x)^2}\right) \ .
\ee
Here $N_{tot}$ is the total number of photons in the PSF. 
We find that eq.~\ref{eq-smu_asym} does an excellent job of 
describing the precision of our simulated $\Delta x$ measurements
for a variety of model PSFs, although in some cases, it may be
necessary to multiply eq.~\ref{eq-smu_asym} by a fudge
factor (to account for the inadequacies of our Gaussian
approximation).

\begin{figure}
\plotone{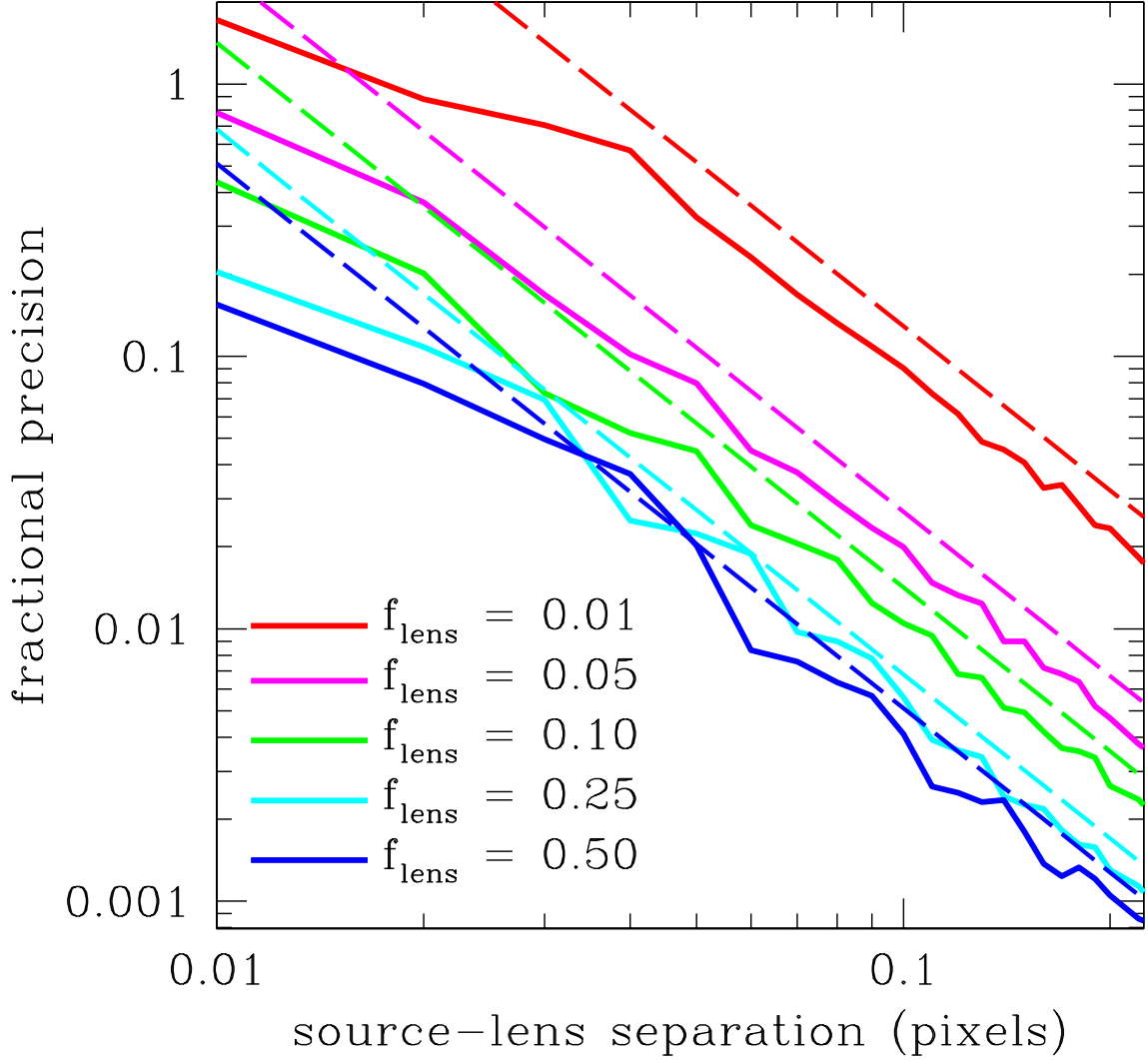}
\caption{
The uncertainty in the lens-source separation,
$\Delta x$, is shown as a function of $\Delta x$. The
different colors indicate different fractions, $f_L$ of the 
blended lens plus source light that is due to the lens.
Solid lines indicate the results of our simulations, and
dashed lines are calculated from eq.~\ref{eq-smu_asym}.
These estimates assume a total of $10^8$ detected
photons from the blended lens plus source.
\label{fig-pm_sig}}
\end{figure}

This approximate formula is compared to the actual 
uncertainties, derived from our simulations in Figure~\ref{fig-pm_sig}
for parameters appropriate for a space-based microlensing
survey, such as the proposed Microlensing Planet Finder
(MPF) mission \citep{mpf-spie}. (Such a mission would have a
PSF FWHM 2-3 times worse than HST, but would have the 
ability to combine images with a combined exposure time of
months.)

We expect that eq.~\ref{eq-smu_asym} may fail to describe the 
measurement uncertainty for $\Delta x$ when the lens-source
separation grows to $\Delta x \simgt s_0$ and the lens and source
images begin to be separately resolved in the high resolution images.
In this case, there are more features of the image that can aid in the
determination of $\Delta x$, 
so we should be able to measure $\Delta x$ even better than 
eq.~\ref{eq-smu_asym} predicts.
However, this is only likely to
occur in cases where $\s{\Delta x}$ is already quite small, and so
eq.~\ref{eq-smu_asym} will simply give a very conservative estimate
of $\s{\Delta x}$ in these cases.

For most planetary events, the light curve shows finite source
effects that allow $t_\ast$ and therefore $\mu_{\rm rel}$ and
$\Delta x$ to be determined from the light curve. 
In these cases, the image elongation
can be used to measure $f_L$, and the uncertainty
in the $f_L$ determination is given by
\be\label{eq-sigma_f}
\s{f_L} = \sqrt{\frac{2}{N_{tot}}} 
\left(\frac{s_0}{\Delta x}\right)^2 \frac{1}{| 1-2f_L|} \ .
\ee
In most cases, this determination of $f_L$ will be redundant with the
determination from the light curve fit and the total lens plus source 
flux. However, if the source star has a bright binary companion,
at a separation $> 10\,$ so that the light curve is not affected,
then these two methods will yield different results.
The binary companion will contribute to the 
$F_S = F_{\rm tot}(1-f_L)$ value
as measured by image elongation, but it will not contribute
to the value of $F_S$ determined from the light curve. Thus,
the different measures of $F_S$ provide
a measurement of the brightness of the source star's binary 
companion.

In order to determine how precisely we can measure both
$\Delta x$ and $f_L$ from the blended image of the source 
plus lens, we will need to measure the third moment,
or skewness, of the blended image profile. The third
moment is given by
\be\label{eq-Qx}
Q = {1\over N_{tot}} \sum (x-x_0)^3 \ ,
\ee
where the sum is over the positions of the detected photons
and $x_0$ is the mean position of the detected photons.
If the image profile is approximately a Gaussian, then
$Q$ can be measured to a precision of 
$\sigma_Q = s^3 \sqrt{15/N_{tot}}$. 
We can also derive the following expression for $Q$ in
terms of $f_L$ and $\Delta x$,
\be\label{eq-Qfdx}
Q = f_L(1-f_L)(1-2f_L) (\Delta x)^3 \ .
\ee
This expression can be used to derive the following 
expressions for the uncertainties in $f_L$ and $\Delta x$
when we solve for both,
\be\label{eq-sigma_dx2}
\frac{\s{\Delta x}}{\Delta x} \approx \sqrt{15\over N_{tot}} {1\over f_L} 
\left({s\over \Delta x}\right)^3 \ ,
\ee
and
\be\label{eq-sigma_f2}
\s{f_L} \approx \sqrt{60\over N_{tot}} \left({s\over \Delta x}\right)^3 \ .
\ee
These expressions assume that $f_L \ll 1$ and that the
uncertainties are dominated by our ability to measure
$Q$, which will generally be the case when $\Delta x/s \simlt 1$.

An additional complication is the possibility of blending
by stars that played no role in the microlensing event,
which happen by chance to lie close to the positions of the source
and lens stars. However, the mean separation of stars in the
Galactic bulge fields that are searched for microlensing events
is in the range of 0.3-1.0 arc seconds, which is much larger
than the lens-source separations. Thus, if the lens-source
star separation can be measured, it is unlikely that stars 
unrelated to the microlensing event will interfere significantly
with the lens-source separation measurement. The only exception
would be stars that are physically associated with the lens 
or source stars, and this possibility is dealt with in \S~\ref{sec-bin}
below. For a space-based microlensing survey, it should generally
be possible to directly detect the unrelated blended stars with
a point-source decomposition algorithm. (This is similar to 
deconvolution \citep{deconvolve} except that it makes use of
the fact these Galactic bulge field contain very few 
sources that are not point-like.)

We should also note that this estimate of our ability
to measure the separation of two unresolved stars
is substantially better than the conservative estimate of 
\citet{han-spaceblend}. Our estimate is justified by our 
detailed simulations, which are, in turn, based on experience with
HST data \citep{andking04}. Thus, we are confident that
the conservative assumptions of \citet{han-spaceblend} are
not needed.

\subsection{Color Dependent Centroid Offset}
\label{sec-color_cent}

Another method to detect the effects of the relative lens-source
proper motion relies upon the likelihood that the lens and
source will have different colors. This means that the lens
star will contribute a different fraction to the total light seen
in the source plus lens image blend as seen in different
passbands. This magnitude of the observed color-dependent
image center shift for the $V$ and $I$ passbands is
\begin{equation}
\Delta x_{V-I} = \left(f_I - f_V\right) \Delta x \ .
\label{eq-dx_VI}
\end{equation}
$\Delta x_{V-I}$ is directly measurable from high resolution
HST images, but $f_I$ and $f_V$ depend
on brightnesses of the lens and source in both passbands.
The brightness of the source star in multiple passbands
can generally be obtained by multicolor observations
at several different magnifications during the course of the
microlensing event, but the brightness of the lens star
is an unknown that must be solved for along with $\Delta x$.
Equations \ref{eq-m_thetaE} and \ref{eq-dx_VI} can be combined
with the $V$ and $I$ band mass-lumonisity relations shown in 
Figure~\ref{fig-mass_lum} to yield four equations for the
four unknowns: $M$, $V_L$, $I_{\rm lens}$, and
$\Delta x = \theta_E T/t_E$. These equations can then be
solved to yield a complete solution for the microlensing event
parameters. This effect has been detected
for the first extrasolar planet discovered by microlensing
\citep{bennett-moa53}, but in this case, $\mu_{rel}$ (and
therefore $\theta_E$ and $\Delta x$) was already known
from the microlensing event light curve.

An important advantage of this method is that for small
$\Delta x$, the signal is $\propto \Delta x$. Thus, as long
as the colors of the source and lens stars differ significantly,
this method should measure small $\Delta x$ vlaues
more precisely than the image elongation method.

\subsection{Interstellar Extinction}
\label{sec-extinct}

Because our methods involve measuring the brightness and
color of the planetary host star in the direction of the Galactic
bulge, interstellar extinction is an important consideration.
Extinction affects the determination of the brightness and
color of both the source and lens stars. The angular
radius of the source star, $\theta_\ast$, is determined
from its brightness and color, and $\theta_\ast$ is used to 
determine the lens-source relative proper motion, $\mu_{\rm rel}$,
and the angular Einstein radius, $\theta_E$. Fortunately,
extinction makes stars appear both fainter and redder, 
and these two effects push our estimate of $\theta_\ast$
in opposite directions. As a result, our estimate of $\theta_\ast$ is
not highly dependent on the uncertainty in our extinction
estimate.

The dust in the Galactic disk is generally modeled with a disk
scale height of $\sim 100\,$pc \citep{drimmel}. Since the
source and lens stars are generally located at a distance
$\simgt 1$-2$\,$kpc, and are $\simgt 2^\circ$ from the
Galactic plane, 
the extinction is primarily in the foreground of both the source 
and lens star. The extinction toward the Galactic bulge is also 
quite patchy, so it cannot be reliably estimated by a simple
model. Instead, the standard practice for the interpretation
of microlensing events is the estimate the extinction toward
the source based
upon the observed colors and magnitudes of stars within
1-2 arc minutes of the microlensed source
\citep{bond-moa53,ogle71,ogle390,ogle169}. For the
lens star, the extinction must be modeled with a 
probability distribution, 
which matches the measured
extinction at the distance of the source and follows the
exponential scaling of \citet{drimmel}. The variance of
the probability distribution accounts for the patchiness of
the actual distribution. In practice \citep{bennett-moa53}, 
this procedure amounts
to only a slight modification to the simpler case in which
all the extinction is assumed to lie in the foreground of both
the lens and source. For simplicity, in this paper, we will assume that
extinction for the lens is identical to the extinction for
the source star, and we will not explicitly discuss the
extinction in the remainder of this paper.

\subsection{Binary Companions to the Lens or Source}
\label{sec-bin}

Although binary stars represent a minority of star systems
\citep{lada_single}, it is important to consider the
effect of a binary source or lens star on our analysis
of the planetary host star properties.
A binary companion to the lens or source star would
generally be unresolved from the source, so it would
complicate our analysis. But, such a possibility is also 
constrained by the properties of the microlensing light curve.

A binary companion to the source star can often be detected via
light curve oscillations due to orbital motion of the source star
if the orbital period is less than a year \citep{eros_spiralarm,96lmc2}.
Microlensing events with main sequence source stars and
detected planetary light curve features generally have a
peak magnification of $\Amax \simgt 10$ because lower 
magnification events typically have poor photometry due
to the crowding in ground-based images. If the event is
well sampled near peak magnification, binary source effects
in the light curve will be visible with an amplitude of $\sim a/(R_E \Amax)$,
where $a$ is the semi-major axis of the source star orbit. For $\Amax > 10$,
this implies that even the orbital motion of a contact binary will be visible
if the photometric precision is close to 1\%. Thus, in most cases, a binary
companion to the source must have a semi-major axis $\simgt 1\,$AU in
order to avoid visible light curve effects. If the source star's companion
is not much fainter than the source, then the lensing of the companion
star may be visible in the light curve unless the source-companion
separation is $\simgt R_E \sim$ several AU. If the separation of the
source from its companion is $\simgt 5$-$10\,$mas, depending on its
brightness, then we will be able distinguish its position from that of the
source star using the same methods that we have discussed to detect
the lens-source offset. For a typical bulge source distance of $9\,$kpc,
this corresponds to a source-companion separation of 45-$90\,$AU, so
for separations larger than this, the binary companion will affect
our measurements of lens-source astrometry.

The light curve constraints on a possible companion to the lens and
planetary host star depend more sensitively on the parameters of the
specific microlensing event. For a very high magnification event,
such as OGLE-2005-BLG-169 (see \S~\ref{sec-ob169}), a companion
of similar mass to the planetary host star can be excluded over a very
wide range of separations from $\sim 0.01$-$100R_E$ or
$\sim 0.03$-$300\,$AU. For events of more modest magnification,
the binary separations are excluded over a more modest range of 
separations, $\sim 0.1$-$10R_E$. So, there will usually be at least
a slim chance of the lens star system is a close binary star 
system with a semi-major axis $a \ll 1\,$AU or a wide binary
system with $a \gg 1\,$AU. The observational signature of
a wide binary companion is similar to the signature of a 
binary companion to the source in that there is an extra source of
light located very close to the positions of the lens and source. 
However, this extra source will move with the planetary host
star, and so the image elongation signal will be stronger.
In contrast, if the planetary host star is a close binary, then
both stars will contribute the lens mass in eq.~\ref{eq-m_rep},
so the lens star system will be fainter and redder than implied by the
single-star mass-luminosity functions shown in Figure~\ref{fig-mass_lum}.

\subsection{Full Solution with Combined High Resolution Follow-up Data}
\label{sec-full_sol}

One complication that may tend to interfere with the recovery of the
planetary host star parameters is the possibility that either the
source star or the planetary host star will have a binary companion
star that contributes a significant amount of flux to the blended
image. Such a star would
generally be unresolved from the position of the lens and source
star, and this would hamper our ability to recover the brightness
of the lens star. Generally, such a binary companion to the
source or lens star would have to have a separation $\gg 10\,$AU
from the lens or source to avoid an obvious light curve
signal. For high magnification microlensing events, a binary
companion would have to have an even larger separation
to avoid detection, but for a low magnification event, it 
would be possible to find a planet orbiting a very close binary
star system with no detectable light curve features.
Fortunately, the high resolution follow-up data
provides more than enough constraints to identify these 
binary companions, in most cases.

The critical unknown parameters for a microlensing event
are the lens mass, distance and relative proper motion,
$\mu_{rel}$, and these are constrained by the light curve
parameters $t_E$ and $t_\ast$ (which is usually, but not always,
measured). The light curve also yields the source star brightness,
usually in two different pass bands, but these measurements
can only constrain the lens star parameters in conjunction
with the high resolution images.
With deep, two-color high resolution space-based
follow-up observations, it is possible to measure five additional
parameters: the source plus lens brightness in two colors,
the image elongation in two colors, and the source plus lens
centroid offset between the two colors. In all, we have 8-9
constraints for the 3 microlensing event unknowns, so the 
problem is significantly over-constrained.

These 5-6 additional parameters can be used to constrain
other possibilities, such as a bright binary companion to the lens
or source star. Such a companion would add 1 parameter
(the companion mass) to the problem or possibly three 
parameters if the separation of the companion is large.
We could probably constrain a companion to the lens, source
or both with these 5-6 additional parameters. However, the
best way to solve a lens system that involves a bright, distant
lens or source companion would be to obtain high resolution
space-based images at multiple epochs. This would add four
additional measured parameters and allow us to distinguish
stellar separation due to the lens-source relative 
proper motion from the separation of a distant binary companion
to the lens or source.

Of course, we haven't
shown that these parameters will always be measured to
sufficient precision to guarantee a complete lens solution.
A complete investigation of this question is beyond the scope
of the present paper, but we will demonstrate that the
question of binaries is easily resolved for the example
event that we discuss in \S~\ref{sec-ob169}.
Also, because $\Delta x$ grows with time, the precision of
our measurements will grow with time, so it is likely that
many ambiguities can be resolved with additional follow-up
observations.
Thus, we can generally expect a complete solution of the
microlensing event when light from the lens can be observed
by HST or another high angular resolution spaced-based telescope.

In practice, some of these measurements can have significant
or correlated uncertainties, so the observational constraints
are best applied through a Bayesian analysis. In fact, such
an analysis is able to provide relatively precise event
parameters even without even observational constraints
to constrain binary companions to the source and lens,
because the {\it a priori} probability of a binary companion
of similar brightness to the source is relatively small. For
example, in the
case of the first microlensing planet discovery
\citet{bennett-moa53} were able to constrain the host star
mass to an accuracy of about $13\,$\%.

\section{A Low-Mass Planetary Event with a Detectable Lens Star}
\label{sec-ob169}

\begin{figure}
\plotone{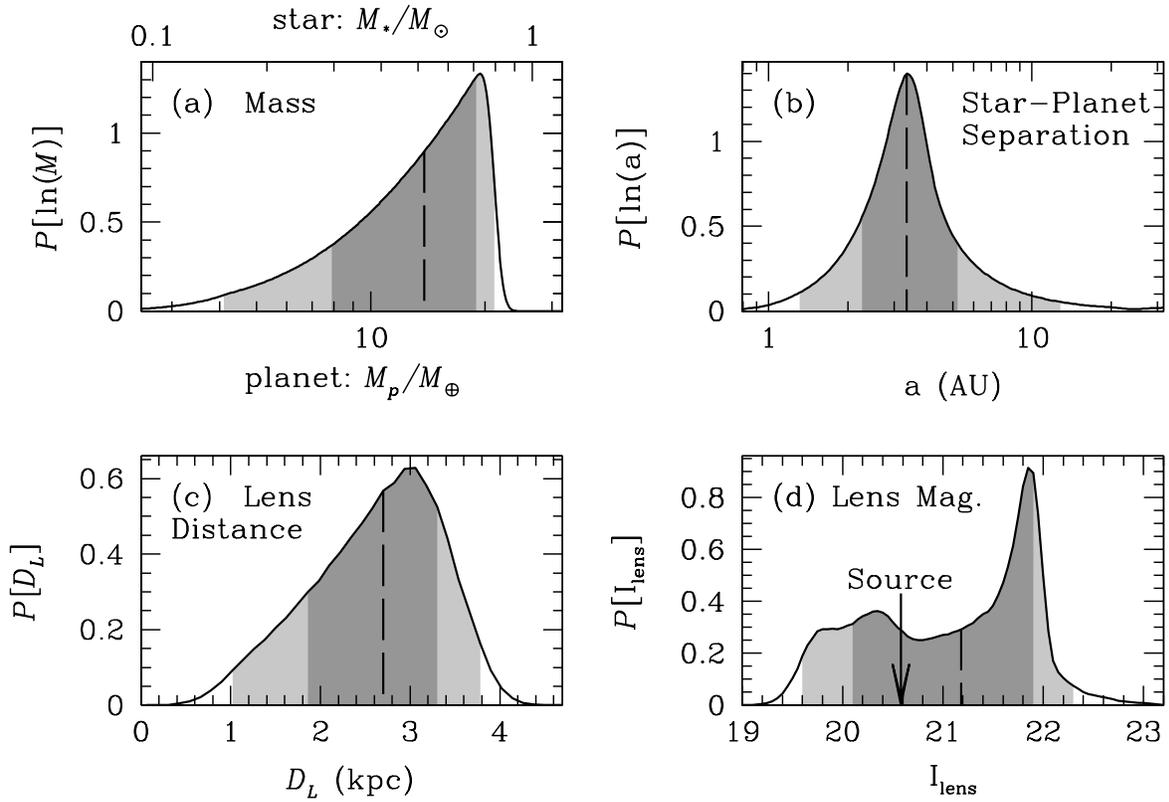}
\caption{
Bayesian probability densities for the properties of the planet
and its host star if it is a main sequence star. 
(a) The masses of the lens star and its planet ($M_\ast$
and $M_p$ respectively). 
(b) the separation, 
(c) their distance from the observer ($D_L$); 
and (d) the I-band brightness of the host star. 
The dashed vertical lines indicate the medians, and the shading
indicates the central 68.3\% and 95.4\% confidence intervals.  All
estimates follow from a Bayesian analysis assuming a standard model for the
disk and bulge population of the Milky Way, the stellar mass function of
Bennett \& Rhie (2002).
\label{fig-lens_prop}}
\end{figure}

Three of the four planets discovered by microlensing to date
had main sequence source stars, which enables detection
of the lens star with HST images a few years after the event.
Of these three planets, OGLE-2005-BLG-169Lb is the most interesting,
since it is a cool, Super-Earth mass planet with
$M_p = 13 {+6\atop -8} \mearth$. Such a planet would be
invisible to other planet
detection methods, and the light curve analysis indicates that it
does not have a gas giant companion in the separation range 1-$10\,$AU. 
Statistical arguments indicate that such 
cool, Super-Earths are the most common type of extrasolar planet 
yet to be discovered \citep{ogle169}.
However the detailed properties of this planet and its host star
are uncertain because the host star has not been detected. In the
absence of the host star detection, the probability distributions for
the star and planet masses, distance, and separation can only
be determined by a Bayesian analysis. This analysis makes 
use of the parameters from the microlensing light curve, including
the lens-source relative proper motion of 
$\mu_{\rm rel} = 8.4\pm 0.6\,{\rm mas/yr}$, and the results are
shown in Figure~\ref{fig-lens_prop}. These have
assumed a Han-Gould model for the Galactic bar
\citep{hangould-mpar}, a double-exponential disk with a
scale height of $325\,$pc, and a scale length of $3.5\,$kpc,
as well as other Galactic model parameters as described in
\citet{gest-sim}. Because this model is slightly different from the
Galactic model used by \citet{ogle169}, the resulting parameters
differ slightly from their results. We find a lens
system distance of $D_L = 2.7{+0.6\atop -0.9}\,$kpc, a three
dimensional star-planet separation of $a = 3.3{+1.9\atop -0.9}\,$AU
and main sequence stellar and planetary masses of 
$M_\ast = 0.52 {+0.19\atop -0.22}\msun$ and 
$M_p = 14 {+5\atop -6}\mearth$. If we assume that white dwarfs
have an {\it a priori} probability to host planets that is equal to
that of main sequence stars (at the separations probed by 
microlensing), then there is a 35\% probability that the host 
star is a white dwarf. The possibility of a brown dwarf host star
is excluded by the light curve limits on the microlensing parallax
effect \citep{ogle169}.

\begin{figure}
\plotone{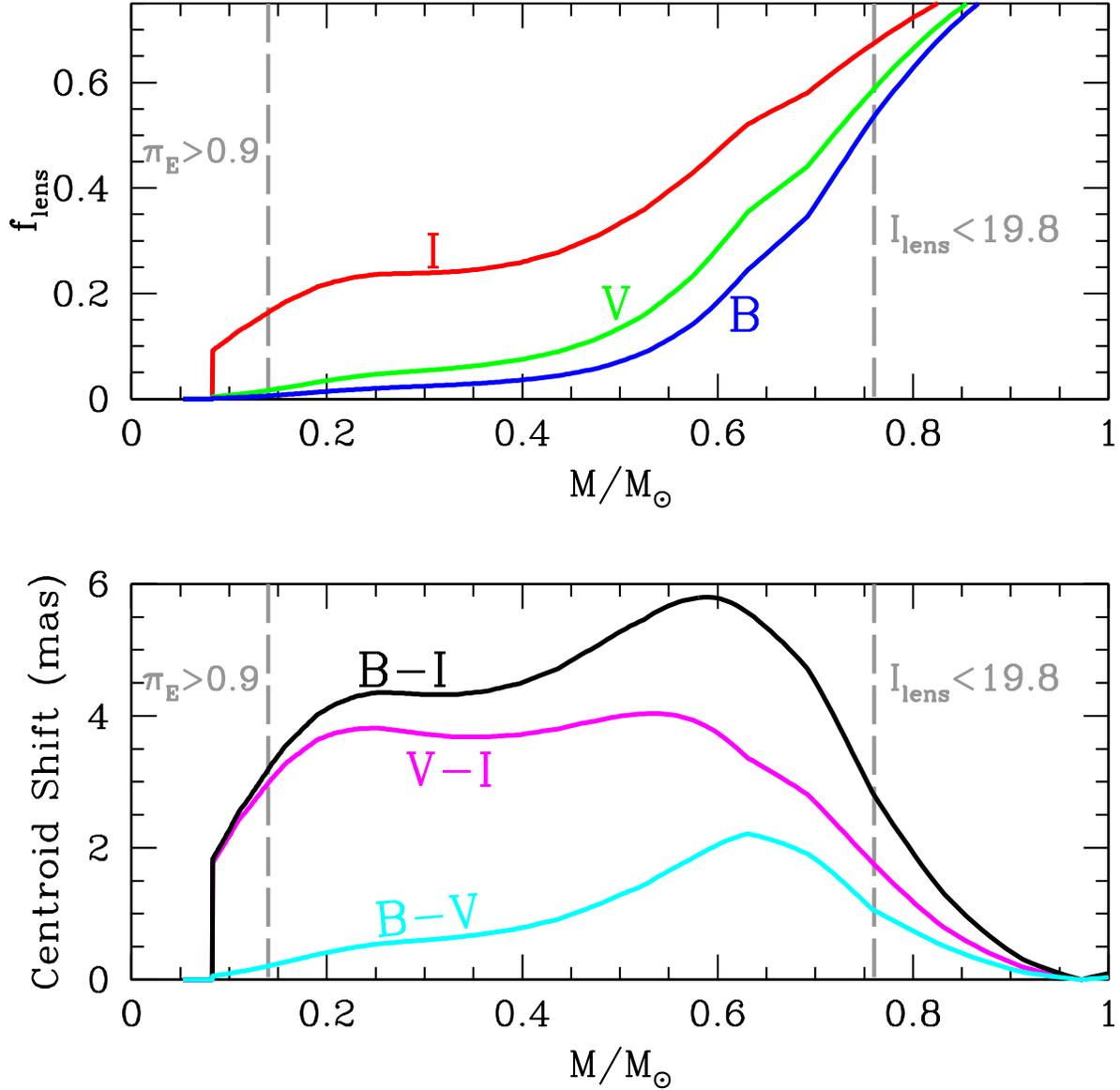}
\caption{
The top panel shows the fraction of the source$+$lens flux that is predicted to
come from the lens in the I (F814W), V (F555W), and B (F435W)
passbands as a function of lens mass. The bottom panel
shows the predicted color-dependent centroid shifts as a function
of mass for 2.4 years of relative proper motion at $\mu_{\rm rel} = 8.4\,$mas/yr.
The grey dashed lines indicate the upper and lower limits on the lens mass
due to the upper limits on the microlensing parallax and lens brightness.
\label{fig-cen_shift}}
\end{figure}

Figure~\ref{fig-lens_prop}(d) shows the probability distribution
of the $I$-band magnitude of the planetary host star compared to the
source star at $I = 20.58\pm 0.10$. The implied planetary host
star brightness distribution has a median and 1-$\sigma$ range of
$I_{\rm lens} = 21.9 {+0.7\atop -1.1}$, but the most interesting
feature of this figure is that the probability of a main sequence
lens fainter than $I = 23$ vanishes. This is because the 
mass-distance relation, eq.~\ref{eq-m_thetaE} ensures that the
lens star will be nearby and at least at bright as $I=23$, even if
it is at the bottom of the main sequence at $M_\ast = 0.08\msun$.
In fact, the microlensing parallax constraint from the light curve
yields a lower limit for the lens star mass of $M_\ast \simgt 0.14\msun$.
Thus, as the upper panel of Figure~\ref{fig-cen_shift} indicates,
the planetary host star must be at least 16\% of the brightness of
the combined lens plus source star blended image, and this implies 
that it will be detectable if it is not a stellar remnant. So, for the faintest
possible main
sequence lens star of the minimum mass, $M_\ast = 0.14\msun$,
eq.~\ref{eq-sigma_f} implies that $f_L$ can be determined with
a precision of 0.017 in a single orbit of HST ACS/HRC observations
in the F814W pass band (with 97,000 detected photons).
This allows a tight constraint on a bright
binary companion to the source star. The situation is a bit
worse for $f_L \sim 0.5$ when the source and lens are almost
equally bright, but in this case, Figure~\ref{fig-cen_shift} 
shows that the lens star will be easily detectable in the $V$ and
$B$ passbands, where the HST PSF ($s_0$) is considerably sharper.
So, we don't anticipate a significant problem in measuring $f_L$,
as long as we have a measurement of $F_S$ in more than one
passband.

If we assume the most likely
case of no detectable companion to the source, then
eq.~\ref{eq-smu_asym} implies that $\mu_{\rm rel}$ can be measured
with a precision of 4.3\% in a single orbit of HST ACS/HRC observations
in the F814W pass band (with 97,000 detected photons). This is an 
improvement over the 7\% measurement of $\mu_{\rm rel}$ that 
comes from the microlensing light curve, and would provide 
independent confirmation of the planetary interpretation of the 
light curve.

The color dependent effects visible in high resolution images of the
OGLE-2005-BLG-169 lens and source star blend are summarized 
in Figure~\ref{fig-cen_shift}, which is based upon the measured
relative lens-source proper motion for this event. 
The top panel shows the fraction 
$f_L$ of  the source plus lens light that is contributed 
by the lens for different HST ACS/HRC passbands, and the
bottom panel shows the predicted color dependent image 
center shifts. Both of these are shown as a function of the planetary
host star mass, and the dashed grey lines indicate the constraints
on the host star mass, $0.14 \msun \leq M_\ast \leq 0.76 \msun$,
implied by the existing observational limits on the microlensing 
parallax effect and the maximum brightness of the lens star.

Several features are apparent in Figure~\ref{fig-cen_shift}. First,
a measurement of a color dependent centroid shift does not
yield a unique mass, because there are usually two different
masses that yield the same centroid shift. This occurs because the
centroid shift can become small if the lens star is faint or if the
lens star has a similar color to the source star. However, this
will generally not create an ambiguity in the interpretation of events
because the brightness of the source star is known. The ``faint lens"
solution implies a fainter total brightness for the lens plus source
blend than the ``similar color" solution. Thus, the centroid shift degeneracy
is broken by the constraint on the lens star brightness, which
we get from the light curve. Also, if the
system is observed in more than two passbands, the 
multiple color dependent centroid shifts will also resolve this degeneracy.

Figure~\ref{fig-cen_shift} also indicates that it will be difficult to 
determine the OGLE-2005-BLG-169L lens star mass precisely if
it is in the range $0.2\msun \simlt M_\ast \simlt 0.4\msun$, because
in this mass range the increase in the $I$-band lens brightness
with distance due to the mass-distance relation (eq.~\ref{eq-m_thetaE})
is nearly compensated for by the decrease in brightness
due to the greater lens distance. This
effect is also responsible for the peak at $I_{\rm lens} \simeq 21.9$
in Figure~\ref{fig-lens_prop}(d). This ambiguity can be resolved
with observations in other passbands. The $V$ and $B$-band
lens brightness fractions and the $B-V$ centroid shift
curve does not have this ambiguity. However, the lens is also 
predicted to be rather faint in these bands if the lens mass is in
the ambiguous range. So, it would be sensible to attack this 
question with a later set of observations if the lens star appears
to lie in the $0.2\msun \simlt M_\ast \simlt 0.4\msun$ mass range
to take advantage of the larger lens-source separation at later times.
Note that
this near degeneracy does not occur for planetary host stars that reside in
the Galactic bulge because the mass-distance relation becomes
quite steep when $D_L$ approaches $D_S$.

\section{Complete Lens Solutions from a Space-Based Microlensing Survey}
\label{sec-mpf}

While follow-up observations with HST or JWST may be feasible for a 
handful of planets discovered each year by microlensing, it will 
probably be difficult to obtain enough observing time to follow-up
all the planets discovered by microlensing when the discovery
rate increases. Certainly, with the expected discovery rate of many
hundreds or even a thousand planets per year from a dedicated
space mission \citep{gest-sim}, like the proposed Microlensing
Planet Finder (MPF) \citep{mpf-spie}, it seems virtually certain that
only a very small fraction of events could be followed up with
observations with HST or JWST. Fortunately, a space-based
microlensing survey will have the capability to perform its own
follow-up observations.

The angular resolution of a dedicated space-based microlensing
survey, like MPF,  will be worse than HST's resolution due to
its smaller aperture (1.1m vs. 2.4m) and near-IR passband. 
However, the decreased angular resolution is more than compensated
for by the much greater total exposure time that will be provided
by a dedicated space telescope. While the MPF PSF may be
three times the width of the HST PSF, but the number of detected
photons in a 1-month MPF exposure can be a factor of 3000 times
the number of photons detected in a single HST orbit. For fixed
$f_L$, eq.~\ref{eq-smu_asym} gives
\be\label{eq-smu}
\frac {\s{\mu}}{\mu_{\rm rel}} \propto 
\frac{s_0^2}{\sqrt{2N_{tot}}\Delta x^2} \,
\ee
so we expect that MPF will be able to measure $\mu_{\rm rel}$
about 6 times more precisely than a single orbit HST observation
(for fixed $\Delta x$).

A space-based survey, like MPF, will have several methods
that can be used to obtain the complete solution of the 
microlensing event. These include:
\begin{enumerate}
\item Combining the mass-distance relation, eq.~\ref{eq-m_thetaE}
with a main sequence mass-luminosity relation similar to those
shown in Figure~\ref{fig-mass_lum}, as described in 
Section \S~\ref{sec-full_sol}, usually gives the most precise results.
The angular Einstein radius, $\theta_E$ can be determined from
the measurement of finite source effects in the light curve
(\ie by determining $t_\ast$), or by a measurement of the 
image elongation, as described in \S~\ref{sec-distort}.
\item The masses and distances can also be estimated from 
the brightness and color measurements of the lens light as
seen by the space-based survey. This method is not generally
as precise as method 1, but it can be considered to be an independent
check on method 1. It can also constrain the possibility of a 
binary companion to the lens star that is too close or distant
for its lensing effects to be apparent \citep{han-spaceblend1},
as discussed in \S~\ref{sec-bin} and \S~\ref{sec-full_sol}.
\item The lens masses can be determined by combining 
measurements of $\mu_{\rm rel}$ and the microlensing 
parallax effect. This has the advantage that it can be used 
even when the lens star is too faint to detect, but a complete 
microlensing parallax measurement will only be available 
for events that are longer than average.
\end{enumerate}

The two main methods for determining the lens-source 
relative proper motion, $\mu_{\rm rel}$, differ
in that the measurement of the source radius crossing 
time, $t_\ast$, only determines the
magnitude of $\mu_{\rm rel}$, while the measurement of
the image elongation determines the full two dimensional
vector, ${\mathbold \mu_{\rm rel}}$. This is useful because it
is generally much easier to measure only a single
component \citep{gould_par_asym}
of the projected Einstein radius, ${\bf \rep}$,
which is a two-dimensional vector with a length
equal to the Einstein radius projected to the position
of the Sun and a direction parallel to the transverse
component of the lens-source relative velocity.
However, since
${\bf \rep} \parallel {\mathbold \mu_{\rm rel}}$, the measurement of
a single component of ${\bf \rep}$ can be combined with
the direction of ${\mathbold \mu_{\rm rel}}$, to yield the 
magnitude of the projected Einstein radius, which is
needed to determine the lens system masses via
eq.~\ref{eq-m_rep}.

\begin{figure}
\plotone{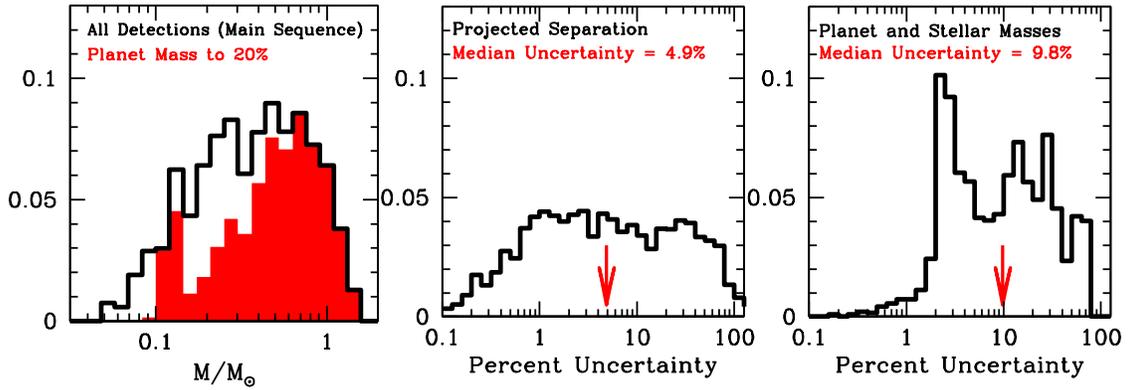}
\caption{
(a) The simulated distribution of stellar masses for stars with 
detected terrestrial planets. The red histogram indicates the
subset of this distribution for which the masses can be determined 
to better than 20\%.
(b) The distribution of uncertainties in the projected star-planet separation.
(c) The distribution of uncertainties in the star and planet masses. Note that
it is the two-dimensional projected separation that is presented here, and we
have not included the uncertainty in the separation along the line-of-sight
as was done in Figure~\ref{fig-lens_prop}.
\label{fig-hist_prop}}
\end{figure}

We have simulated \citep{gest-sim} the planetary events
expected for a space-based microlensing survey similar
to the MPF mission \citep{mpf-spie} to determine how
well the parameter of the detected planetary systems
can be determined. The results of these simulations
are displayed in Figure~\ref{fig-hist_prop}. The parameters
are solved for using methods 1 and 3 above, and the
method that gives the most precise parameters is reported
in Figure~\ref{fig-hist_prop}. One simplifying assumption that
we have made is to assume that only the single component
microlensing parallax asymmetry can be measured, instead
of the full ${\bf \rep}$ vector. This means that we don't have 
a direct method to determine the lens masses for planetary
host stars that are too faint to detect. (These are generally 
white dwarfs, brown dwarfs, and late M-dwarfs). If we assume 
that planetary masses scale in proportion to the host star mass, 
but does not otherwise depend on stelar type, then about
24\% of stars would fall into this ``invisible star" category.

Figure~\ref{fig-hist_prop} considers only planets found orbiting
main sequence host stars, and panel (a) indicates that 
the distribution of host star masses is rather flat with
precise mass determinations most common for the
more massive host stars. Panels (b) and (c) show the
distributions of the separation and mass uncertainties
from these simulations with median uncertainties of
5.2\% and 10.2\%, respectively. These simulations clearly
show that for the majority of planets discovered by a space-based
microlensing survey, the star and planet masses, separation
and host star type will be determined with reasonable
precision.

\section{Discussion and Conclusions}
\label{sec-conclude}

We have shown that the main uncertainties in the properties of
planetary microlensing events can be overcome with
space-based observations. For planets detected with
ground-based microlensing observations, space-based
follow-up imaging can detect the lens star, which allows
the properties of the star and its planet to be determined.
For a space-based microlensing survey, no follow-up observations
are needed because the survey data will contain the information
needed to identify the lens star and solve for the masses and
separation.

The space-based observations in this paper apply primarily to
events with main sequence source stars. Main sequence stars
are the prime targets for
space-based microlensing surveys \citep{gest-sim} and for
ground-based attempts to find earth-mass planets
\citep{em_planet,wamb}. However, giant source stars do allow the
detection of planets a bit more massive than the Earth, including the
lowest mass planet detected to date, OGLE-2005-BLG-390Lb,
\citep{ogle390}, around a main sequence star. (Of course,
much lower mass planets have been discovered orbiting
a pulsar \citep{pulsar_planets,pulsar_planets03}.)

While giant source stars are attractive targets for microlensing
planet searches because their photometry is less affected by
blending and because they require much shorter exposure times
for precise photometry, they have the drawback that the source
star is generally brighter than the lens star by a large factor.
In eq.~\ref{eq-smu_asym}, if $F_s \gg F_l$,
then $N_{tot} \propto F_s$ and $f_L \propto 1/F_s$,
so $\s{\mu}/\mu_{\rm rel} \propto 1/\sqrt{F_s}$. Thus, since
typical red clump giant sources are brighter than the typical
main sequence source stars by a factor of $\sim 100$, we
expect that $\mu_{\rm rel} $ could be measured about ten
times worse for a giant source than for the same event parameters
with a main sequence source star. It would 100 times more exposure
time to compensate for this, which seems implausible in all but
the most favorable cases. A more reasonable strategy would be to
wait longer to allow $\Delta x$ to grow, or to switch to an instrument
with much higher angular resolution than HST, 
such as the Very Large Telescope Interferometer (VLTI)
\citep{vlti2006}. While current VLTI instrumentation is unable to
observe the lens-source separation \citep{vlti},
it is expected that a future VLTI instrument will be
capable of detecting the OGLE-2005-BLG-390L
lens star sometime in the next decade (Beaulieu \etal, in preparation).
It is also possible that high resolution adaptive optics imaging could be used to
make some of the measurements that we have described here
for events with main sequence source stars. This would require
either very precise knowledge of the AO imaging PSF, or
a longer delay between the planetary event and the follow-up
imaging to provide a larger lens-source separation.

Our investigation has been incomplete in that we have only considered
a subset of the information that might come from analysis of the
microlensing parallax effect. For a small subset of very long timescale
events, it is likely that the lens system parameters can be characterized
much more precisely than we have estimated here. It is also likely
that many events with a parallax asymmetry measurement, but without
a measurement of the two-dimensional ${\bf \rep}$ vector will be
able to have their parameters measured determined due to future
high precision measurements of
${\mathbold \mu_{\rm rel}} \parallel {\bf \rep}$
with future high angular resolution instruments like the VLTI or
the James Webb Space Telescope \citep{jwst} with the added
benefit of a long time baseline.

This paper has focused on the situation in which signals of both
the lens star and planet are detected, but it is also possible to detect
planets that are so far from their parent star that the star does not
yield a photometric microlensing signal
\citep{distefano-widesep,han-widesep}. In such a situation, it
may be difficult to distinguish the light curves of planets in
very distant orbits from those of unbound planets that have been
ejected from the planetary systems that they were born in. But in
the former case, the planetary host star will often be detectable
by the methods described in this paper. If these methods should
fail to detect the host star, follow-up observations with the 
Space Interferometry Mission (SIM) will allow the host star to
be discovered via its astrometric gravitational microlensing
signal \citep{han-freefloat}, which will work even for dark host
stars, such as white or brown dwarfs. However, SIM observations
of main sequence source stars may be limited by the long
exposure times required for sources fainter than $V = 19$.

It is often said that microlensing planet discoveries cannot be
followed up, but this is not completely true. While a repeat
of the photometric planetary signal will not usually occur for
another $\sim 10^6\,$years, we have shown that
most planetary events provide a prediction of the lens-source 
relative proper motion, $\mu_{\rm rel}$, which can be confirmed with
high angular resolution follow-up observations with HST.
Furthermore, these follow-up observations also allow the 
complete solution of the microlensing event, which includes
the conversion of the planetary mass ratio into physical
masses for the planet and its host star, as well as the projected
star-planet separation in physical units. Finally, we have also 
shown that a dedicated space-based microlensing survey,
such as MPF, will collect the data necessary to extract these
follow-up observations in its own photometric survey images.
Thus, a space-based microlensing survey will provide
full lens event solutions, with planet and star masses, their
separation in physical units, and their distance from us, for
most of the planets that are discovered.

\acknowledgments

D.P.B. was supported by
grants AST 02-06189 from the NSF and NAG5-13042 from
NASA. J. A. was supported by NASA/HST Grant GO-9443.

\clearpage



\clearpage

\end{document}